\documentclass[ps,reprint,floatfix,superscriptaddress,twocolumn]{revtex4-2}
\usepackage[english]{babel}
\usepackage{amsmath,blkarray,amsfonts}
\usepackage{physics}
\usepackage{epsfig}

\begin{document}
\title{Kinetic equation from Landau level basis: Beyond relaxation-time approximation}
\author{Kitinan Pongsangangan}
\email{kitinan.pon@mahidol.ac.th}
\affiliation{Department of Physics, Faculty of Science, Mahidol University, Bangkok 10400, Thailand}
\begin{abstract}
The purpose of this paper is to formulate a kinetic theory describing transport properties of electrons in a uniform magnetic field of arbitrary magnitude.  Exposing an electronic system to a constant magnetic field quenches its energy bands into a series of discrete energy levels, known as Landau levels. The Landau-level states, exact solutions of the Schr\"odinger equation in a constant background magnetic field, are natural and suitable basis to use, especially, for the investigation of strong-magnetic-field phenomena. Starting from the Keldysh formalism, we derive the quantum kinetic equation from the Landau-level basis. As an illustration, we apply the kinetic equation to calculate the electrical conductivity of a two-dimensional electron gas exposed to a perpendicular magnetic field.
\end{abstract}
\maketitle
\section{Introduction}
Boltzmann equation is one of the standard frameworks for transport phenomena in condensed matter systems. For the Boltzmann equation to be valid, the existence of well-defined quasiparticle excitations is required. The quasiparticle spectral function relates energy to momentum via a dispersion relation allowing for defining a distribution function at a position $\vec{r}$ at time $t$ in a momentum state $\vec{p}$.  The Boltzmann equation governs its time evolution under the influence of external disturbances, such as, a temperature gradient and an electric field. In several circumstances, this description ceases to be valid. An example of particular interest for this work is the quantum Hall effect.  

Applying a uniform magnetic field perpendicular to a two-dimensional electronic system quenches its energy bands into a series of discrete Landau levels. Such a singular system renders an ideal platform for studying quantum transport phenomena beyond the semiclassical Boltzmann description. Momentum is no longer a good quantum number. The Landau levels may be broadened substantially due to interactions \cite{tsuneya1974I,tsuneya1974IV,MacDonald1991}. The spectral function, in this case, deviates significantly from Dirac delta peaks.  In order to take the effect of level broadening into account, a quantum analog of the distribution function which depends independently on energy variable is called for.  The Landau-level index is needed to be incorporated. 

Several attempts in formulating a kinetic equation for electrons in the Landau-level state have been pushed forward for both relativistic and non-relativistic systems \cite{lifshits1958a,lifshitz1958,silin1958,Argyres1958,Argyres1960,andreevkosevich1961,Levinson1969,Levanda1994,Laikhtman1994,Aleiner2004,Gorbar2017,Wang2018,Al-Naseri2020,Lixin2020}. An open question of particular interest for this work is how to systematically incorporate collision integrals for various scattering processes. The relaxation-time approximation is often employed for the electron-impurity scattering \cite{Gorbar2017}. The scattering time is frequently assumed to be magnetic-field independent. In many cases, this assumption is insufficient even for weak magnetic fields. It was suggested, for instance, that, within the self-consistent Born approximation (SCBA), the electron-impurity self-energy depends significantly on an applied magnetic field \cite{Mahan1984,Mahan1987,tsuneya1974IV,Mirlin2015,Mirlin2017}.  The other scattering mechanisms, for example, due to electron-phonon \cite{MacDonald1991,Pound2011} and electron-electron interactions \cite{silin1958} have been scarcely discussed in literatures. In this work, we attempt to fill this gap by addressing how to derive a kinetic equation with collision integral for an electronic system subjected to a strong magnetic field. Our starting point is the Keldysh formalism.

The Landau-level wave-function is the exact solution of the Schr\"odinger equation with a constant background magnetic field. It forms a natural and suitable basis to use, especially, for the investigation of strong-magnetic-field phenomena. We develop a quantum kinetic equation for transport properties in this limit by using the Landau-level basis and the Keldysh-field-theoretical technique \cite{KadanoffBaym1962,Keldysh1965,Kamenev2011,CalzettaHu2008,rammer2007}. Similar to the other quantum field theories, one of the advantages of Keldysh formalism over the other formalisms is that  several tools, such as, perturbative expansion and Feynman diagrammatic technique, are available for a systematic calculation of self-energies and, consequently, collision integrals. The basic objects of the Keldysh formalism are the retarded and the Keldysh Green functions. The former contains the information about the density of state while the latter is the quantum counterpart of the distribution function. They satisfy Dyson equations which are, in practice, tractable with the help of the Wigner transformation and a gradient-expansion approximation. This gives a fully quantum-mechanical generalization of the semi-classical Boltzmann transport theory. 

One issue required a careful consideration is gauge symmetry of the Green functions. In the presence of an electromagnetic field, the fermionic Green functions are not gauge-invariant. To circumvent this issue, Schwinger introduced a phase factor, later named after him. This phase factor was introduced in such a way that it exactly compensates with  the phase coming from fermion fields under gauge transformation \cite{Schwinger1951}. In the same spirit, we modify the Wigner transformation  by including such a phase. This ensures that the Green functions, their inverses as well as the resulting quantum kinetic equation are invariant under gauge transformation. As a side remark, the effect of this modification coincides with the shift of variables introduced by Mahan in \cite{Mahan1984,Mahan1987}. 

As an application of our quantum transport equation, we calculate the electrical conductivity of a 2DEG in the quantum hall regime in the presence of impurities. We find that, as a result of the discreteness of the Landau levels, a solution to our quantum kinetic equation shows the plateaus of Hall conductivity accompanied by a series of peaks in the longitudinal conductivity.

Throughout the paper, we set $e=\hbar=k_B =1$, unless stated otherwise, and restore them in the final results. Here $-e$ is the electric charge of an electron, $\hbar$ is the reduced Planck's constant and $k_B$ is the Boltzmann constant.

 This paper is organized as follows. In Sec.~\ref{sec:Keldyshformalism}, we introduce the Keldysh technique and present the derivation the quantum kinetic equation.  We modified the Wigner transformation by including the Schwinger phase such that the resulting Green functions and the quantum kinetic equation are gauge invariant in the presence of a strong magnetic field. We apply the kinetic equation to calculate electrical conductivity for a two-dimensional electron gas in the quantum Hall regime.  In section \ref{sec:model}, we introduce the model. As for the self-energies, we consider the electron-impurity interaction within the self-consistent Born approximation. Sec.\ref{sec:results} is devoted to the solution of the kinetic equation and the calculation of the conductivities. It is the main result of this work. Conclusion and outlooks are discussed in Sec.\ref{sec:conclusion}.


\section{Formalism: Derivation of the Quantum Kinetic equation}
\label{sec:Keldyshformalism}

\subsection{Keldysh formalism}

This section is devoted to the derivation of the quantum kinetic equation from the non-equilibrium quantum field theory. The basic object of the method is the Green functions casted in a matrix form. Following Larkin and Ovchinnikov convention \cite{LarkinOvchinnikov1975}, the Green function matrix for fermions can be conveniently arranged as
\begin{equation}
\hat{G}(\underline{x}_1,\underline{x}_2) = 
\begin{pmatrix} 
G^R(\underline{x}_1,\underline{x}_2) & G^K(\underline{x}_1,\underline{x}_2)  \\ 0 & G^A(\underline{x}_1,\underline{x}_2)
\end{pmatrix}.
\label{eq:dressedG}
\end{equation}
Hereafter we employ the four-vector notation $\underline{x}=(t,\vec{r})$ for space-time coordinate. 

The retarded and advanced Green functions denoted by the superscripts $R$ and $A$ define the density of state $\rho$ according to 
\begin{equation}
\rho(\underline{x}_1,\underline{x}_2)  = -\frac{1}{2\pi i } \left( G^R(\underline{x}_1,\underline{x}_2) - G^A(\underline{x}_1,\underline{x}_2) \right).
\end{equation}
The Keldysh Green function denoted by the superscript $K$ plays the role of the quantum counterpart of the phase-space distribution function. Typically, it is decomposed into a product of the distribution function and the density of state according to
 \begin{eqnarray}
 G^K(\underline{x}_1,\underline{x}_2) &=& (G^R \circ F - F \circ G^A)(\underline{x}_1,\underline{x}_2),\\
 \end{eqnarray}
The convolution is defined as integrals over intermediate space-time variables as well as summations over matrix indices. Here $F = 1-2f$. In thermal equilibrium, $f$ reduces to the Fermi-Dirac function.
For an interacting theory, the dressed Green functions in Eqs. \eqref{eq:dressedG} satisfy Dyson's equations according to
\begin{equation}
\left( \hat{G}^{-1}_0 - \hat{\Sigma} \right) \circ \hat{G} = \hat{1},
\label{eq:Dysonequation-a}
\end{equation}
\begin{equation}
 \hat{G} \circ \left( \hat{G}^{-1}_0 - \hat{\Sigma} \right) = \hat{1},
 \label{eq:Dysonequation-b}
\end{equation}
 The inverse propagator $\hat{G}_0^{-1}(\underline{x}_1,\underline{x}_2)$ and the self-energy $\hat{\Sigma}(\underline{x}_1,\underline{x}_2)$ of fermions possesses the same matrix structure as that of the Green function. They read as

\begin{equation}
\hat{G}_0^{-1}(\underline{x}_1,\underline{x}_2) = 
\begin{pmatrix} 
[G^{-1}_0]^R(\underline{x}_1,\underline{x}_2) & 0 \\ 0 & [G^{-1}_0]^A(\underline{x}_1,\underline{x}_2)
\end{pmatrix},
\end{equation}

\begin{equation}
\hat{\Sigma}(\underline{x}_1,\underline{x}_2) = 
\begin{pmatrix} 
\Sigma^R(\underline{x}_1,\underline{x}_2) & \Sigma^K(\underline{x}_1,\underline{x}_2)  \\ 0 & \Sigma^A(\underline{x}_1,\underline{x}_2)
\end{pmatrix}.
\end{equation}
Taking the difference of the Dyson's equations, the  Keldysh component gives \begin{equation}
\label{eq:KeldyshequationG}
\left[G^{-1}_0~ \overset{\circ}{,} ~G^{K} \right] = \Sigma^{R} \circ G^K + \Sigma^{K} \circ G^A -G^{R} \circ \Sigma^K -G^{K} \circ \Sigma^A.
\end{equation}
This constitutes the starting point for the derivation of the quantum kinetic equation.
As will be presented in the subsequent subsection, the Wigner transform of these equations together with a gradient-expansion approximation gives a quantum-mechanical generalization of the semi-classical Boltzmann transport theory. These equations are usually referred to as the quantum kinetic equations. 




\subsection{Gauge-invatiant Wigner Transformation}
The Keldysh equations \eqref{eq:KeldyshequationG} is, in general, exceedingly complicated coupled system of equations. However, in the limit in which external disturbances vary slowly in space and time, it usually becomes much simpler and more tractable with the help of the Wigner transformation and the gradient-expansion approximation.  In this section, we review the definition of the Wigner transformation and generalize it for fermions exposed to an external electromagnetic field. 

The Wigner transformation of a two-point function is defined as a Fourier transform with respect to its relative coordinates. For a two-point function $b(\underline{x}_1,\underline{x}_2)$, we introduce a relative $\underline{x} =\underline{x}_1-\underline{x}_2$ and center-of-mass coordinates $\underline{X}=\frac{\underline{x}_1+\underline{x}_2}{2}
$. Its Wigner transform reads as
\begin{equation}
b(\underline{X},\underline{p}) = \int d\underline{x} e^{-i\underline{p} \cdot \underline{x}} b\left( \underline{X},\underline{x}\right).
\end{equation}
 Here we use the combined notation for the dot product of the space-time $\underline{x}=(t,\vec{r})$ and the energy-momentum $\underline{p}=(\omega,\vec{p})$ as  $\underline{p}\cdot\underline{x}= \vec{p}\cdot\vec{r}-\omega t$.  

The first step towards the quantum kinetic equation is to map the Keldysh equations \eqref{eq:KeldyshequationG} into the Wigner space.  Let  $b(\underline{x}_1,\underline{x}_2) = g \circ h \equiv \int d\underline{x}_3~g(\underline{x}_1,\underline{x}_3) ~h(\underline{x}_3,\underline{x}_2)$. The Wigner transform of $b$ known as the Moyal product is given by
\begin{widetext}
\begin{equation}
\label{eq:wignertransformboson}
\begin{split}
\tilde{b}(\underline{X},\underline{p}) = \sum_{\alpha ,\beta=0}^{\infty} \frac{1}{\alpha!\beta!}\left(-\frac{i}{2}\right)^\alpha \left( \frac{i}{2}\right)^\beta \left[ \partial^\alpha_{\underline{X}_{\mu}} \partial^\beta_{\underline{p}_{\nu }}g(\underline{X},\underline{p}) \right] \left[ \partial^\beta_{\underline{X}_{\nu}} \partial^\alpha_{\underline{p}_{\mu }} h(\underline{X},\underline{p}) \right].
\end{split}
\end{equation}
\end{widetext}
Here $h(\underline{X},\underline{p})$ and $g(\underline{X},\underline{p})$ are the Wigner transforms of $h(\underline{x}_1,\underline{x}_2)$ and $g(\underline{x}_1,\underline{x}_2)$, respectively. 

When fermions coupled to an external electromagnetic field, their Green function and the Wigner transform defined above violates gauge symmetry. To overcome this problem, we include the Schwinger phase $\phi(x_1,x_2)$ \cite{Schwinger1951} and redefine the Wigner transform as 
\begin{equation}
\label{eq:wignertransformation}
\tilde{f}(\underline{X},\underline{p}) = \int d\underline{x}  e^{-i\underline{p}\cdot\underline{x}} e^{i \phi(\underline{X},\underline{x})} f(\underline{X},\underline{x}).
\end{equation}
with
\begin{equation}
\label{eq:Schwingerphase}
\phi(\underline{x}_1,\underline{x}_2)=-\int_{\underline{x}_1}^{\underline{x}_2}d\underline{x}' \cdot \underline{A}(\underline{x}') - \frac{1}{2}\int_{\underline{x}_1}^{\underline{x}_2} d\underline{x}'\cdot \underline{\underline{F}} \cdot \left( \underline{x}' - \underline{x}_1\right).
\end{equation}
Here
$\underline{\underline{F}}_{\mu\nu}=\partial_{\underline{x}^\mu}\underline{A}_{\nu}(\underline{x})-\partial_{\underline{x}^\nu}\underline{A}_{\mu}(\underline{x})$ defines the gauge-invariant strength tensor in terms of the four-vector gauge potential $\underline{A}_\mu(\underline{x})$. In what follows, we choose to work with $\underline{A}(\underline{x}) \equiv (A^t,\vec{A})= (0,-Et,Bx,0)$ where the Landau gauge is employed for the magnetic field and a weak external electric field $\vec{E}= -\partial_{\vec{r}^i} \underline{A}_t(\underline{x}) - \partial_{t} \underline{A}_i(\underline{x})  = E\hat{x}$ is applied along the x-axis. In the subsequent sections, we will be interested in a linear response of  a 2DEG in a strong magnetic field background to a weak electric field. It can be shown that the value of the Schwinger phase defined in Eq.\eqref{eq:Schwingerphase} is path-independent. This is assured by the Stoke's theorem as the integrand is curl-free. For simplicity, we choose to evaluate the Schwinger phase on a straight line joining the boundaries of the integration, $\underline{x}_1$ and $\underline{x}_2$.  
We find that
 
\begin{equation}
\begin{split}
\phi(x_1,x_2) \rightarrow \phi(\underline{X},\underline{x}) 
&= -Er^xT+Br^yR^x.
\end{split}
\label{eq:schwingerphase}
\end{equation} 


Consider a two-point function $b(\underline{x}_1,\underline{x}_2) = \int d\underline{x}_3~g(\underline{x}_1,\underline{x}_3) ~h(\underline{x}_3,\underline{x}_2)$. 
With the Schwinger phase, its Wigner transform becomes

\begin{widetext}
\begin{equation}
\begin{split}
\label{eq:genMoyalproduct}
b(\underline{X},\underline{p}) &= \sum_{\substack{\beta \ge \gamma=0 \\ \delta \ge \alpha=0}}^{\infty} \frac{1}{\alpha!\beta!\gamma!\delta!} \sum_{\mu,\nu=0}^{3} \left( -\partial_{\underline{x}^\mu_{a}}\right)^{\delta-\alpha} \left( -\partial_{\underline{x}^\nu_{b}}\right)^{\beta-\gamma} e^{i\phi\left(\underline{X},\underline{x}_a-\underline{x}_b\right) - i \phi\left(\underline{X}+\frac{\underline{x}_a}{2},-\underline{x}_b\right) - i\phi \left( \underline{X}+\frac{\underline{x}_b}{2},\underline{x}_a\right) } \Big|_{\substack{\underline{x}_a=0 \\ \underline{x}_b=0}}  \\ & \hspace{3cm} \left[ \left( -\frac{1}{2}\partial_{\underline{X}^\mu}\right)^\alpha \left( i\partial_{\underline{p}^\nu} \right)^\beta g\left( \underline{X},\underline{p}\right) \right] \left[ \left( -\frac{1}{2}\partial_{\underline{X}^\nu}\right)^\gamma \left( -i\partial_{\underline{p}^\mu}\right)^\delta h \left( \underline{X},\underline{p}\right) \right],
\end{split}
\end{equation}
\end{widetext}
The addition exponential function arises from the Schwinger phase. With the gauge choice specified above, we find that
\begin{equation}
\begin{split}
&i\phi\left(\underline{X},\underline{x}_a-\underline{x}_b\right) - i \phi\left(\underline{X}+\frac{\underline{x}_a}{2},-\underline{x}_b\right) - i\phi \left( \underline{X}+\frac{\underline{x}_b}{2},\underline{x}_a\right) \\ & \hspace{2cm}= -i\frac{E}{2}\left( r^x_{b}t_a-r^x_{a}t_b\right) + i\frac{B}{2}\left( r^y_{b}r^x_{a}-r^y_{a}r^x_{b}\right)
\end{split}
\end{equation} 
As expected, by setting the Schwinger phase to zero, this expression reduces to the Moyal product, Eq.\eqref{eq:wignertransformboson}.

\subsection{Quantum Kinetic Equation}

Let us consider the Keldysh equation \eqref{eq:KeldyshequationG} and transform it to the Wigner space  by making use of Eq.\eqref{eq:genMoyalproduct}. We expand the left-hand side of the resulting equation up to the second order in derivatives ($\alpha+\beta+\gamma+\delta \le 2$). This gives the streaming terms of the quantum kinetic equation reading as
\begin{equation}
\begin{split}
\left[G^{-1}_0~ \overset{\circ}{,} ~G^{K} \right] \xrightarrow{\text{WT}} &~~ i\partial_{\vec{R}} \tilde{G}^{-1}_0 \cdot \partial_{\vec{p}} \tilde{G}^K - i \partial_{\vec{p}} \tilde{G}^{-1}_0 \cdot \partial_{\vec{R}} \tilde{G}^K \\ &-i\partial_{T} \tilde{G}^{-1}_0  \partial_{\omega} \tilde{G}^K + i \partial_{\omega} \tilde{G}^{-1}_0 \partial_{T} \tilde{G}^K
 \\ &+ iE\partial_{p_x}\tilde{G}^{-1}_0 \partial_{\omega}\tilde{G}^K - iE\partial_{\omega}\tilde{G}^{-1}_0 \partial_{p_{x}}\tilde{G}^K \\ &+ iB \partial_{p_y} \tilde{G}^{-1}_0 \partial_{p_{x}} \tilde{G}^K -iB \partial_{p_x} \tilde{G}^{-1}_0 \partial_{p_{y}} \tilde{G}^K.
\end{split}
\end{equation}
Here $\tilde{G}^{-1}_0$ and $\tilde{G}^K$ are the Wigner transforms of $G^{-1}_0$ and $G^K$, respectively. They are functions of the the center-of-mass space-time coordinates $(\vec{R},T)$ as well as momentum and frequency $(\vec{p},\omega)$. For brevity, here, we omit their arguments.

On the other hand, the right-hand side is expanded up to the zero order in gradient. This gives the collision integral. It reads as
\begin{equation}
\begin{split}
& \Sigma^{R} \circ G^K + \Sigma^{K} \circ G^A -G^{R} \circ \Sigma^K -G^{K} \circ \Sigma^A  \\ &~~\xrightarrow{\text{WT}} \left( \tilde{\Sigma}^{R} - \tilde{\Sigma}^{A} \right)  \tilde{G}^K - \tilde{\Sigma}^{K} \left( \tilde{G}^R - \tilde{G}^{A} \right) 
\end{split}
\end{equation}

For brevity, we also omit the arguments of the self-energies $\tilde{\Sigma}^{R/A/K}(\vec{R},T,\vec{p},\omega)$

\section{Model}
\label{sec:model}

\subsection{2DEG}
As an application of the transport equations developed above, we theoretically study transport properties of an electron gas in two dimensions subjected to a perpendicular magnetic field. The Keldysh action for the 2DEG takes the form
\begin{equation}
S_{\text{2DEG}}[\psi^\dagger,\psi] = \int d\underline{x}_1 d\underline{x}_2  \sum_{a,b=1}^{2} \psi^\dagger_{a}(\underline{x}_1)  \hat{G}_{0ab}^{-1}(\underline{x}_1,\underline{x}_2) \psi_{b}(\underline{x}_2).
\end{equation}
Here the inverse Green function is arranged in a matrix as
\begin{equation}
\hat{G}_0^{-1}(\underline{x}_1,\underline{x}_2) = 
\begin{pmatrix} 
[G^{-1}_0]^R(\underline{x}_1,\underline{x}_2) & 0 \\ 0 & [G^{-1}_0]^A(\underline{x}_1,\underline{x}_2)
\end{pmatrix}
\end{equation}
with 
\begin{equation}
\begin{split}
[G^{-1}_0]^{R(A)}(\underline{x}_1,\underline{x}_2) &= \delta(\underline{x}_1-\underline{x}_2) \times\\ &\left( i\partial_{t_2} + \frac{1}{2m} \left(\vec{\nabla}_{\vec{r}_2} + \vec{A}(\underline{x}_2)\right)^2 \pm i0\right).
\end{split}
\label{eq:inversepropoagator}
\end{equation}
We ignore the effect of Zeeman splitting. 

The Schr\"odinger equation for an electron in a constant background magnetic field is solvable exactly by the Landau-level wavefunction.  The solution in the Landau gauge reads as

\begin{equation}
\label{eq:eigenstate}
\psi_{np_y}(\vec{r}) = \frac{i^n B^{1/4}}{\sqrt{2^n n!\sqrt{\pi}}}e^{-\frac{B}{2}\left(x+\frac{p_y}{B}\right)^2} H_n\left(\sqrt{B}\left(x+\frac{p_y}{B}\right)\right) e^{ip_yy},
\end{equation}
with quantized eigen-energies given by
\begin{equation}
\epsilon_n  = \omega_c \left( n + \frac{1}{2}\right) \hspace{0.5cm} \text{for} \hspace{0.2cm} n \ge 0. 
\end{equation}
Here $H_n(x)$ denotes the Hermite polynomial of order $n$ and  $\omega_c=B/m$ defines the cyclotron frequency.

The Landau-level wavefunctions form a natural and suitable basis to expand the bare Green functions:

\begin{equation}
\label{eq:greenfuntionsinLandaubasis}
G_0^{R/A/K}(\vec{r}_1,t_1,\vec{r}_2,t_2) = \sum_{np_y} \psi^*_{np_y}(\vec{r}_1)\psi_{np_y}(\vec{r}_2) G^{R/A/K}_{0np_y}(t_1,t_2)
\end{equation}

The Green functions are diagonal in the Landau-level space. Here

\begin{equation}
\begin{split}
G^{R/A}_{0np_y}(\nu) &= \frac{1}{\nu-\epsilon_n \pm i0^+},  \\
G^{K}_{0np_y}(\nu) &= \left( G^{R}_{0np_y}(\nu)-G^{A}_{0np_y}(\nu)\right) \left( 1-2f(\nu)\right),
\end{split}
\end{equation}
with $f(\nu)$ being the Fermi-Dirac distribution function in thermal equilibrium.

We find that their Wigner transforms, according to Eq.\eqref{eq:wignertransformation} are given by 
\begin{equation}
\label{eq:GreenfunctioninLandaulevelbasis}
\tilde{G}_0^{R/A/K}(\vec{p},\nu)  =  \sum_{n}  e^{-\frac{p^2}{B}} 2(-1)^n  L_n\left( \frac{2p^2}{B}\right) G^{R/A/K}_{0n}(\nu).
\end{equation}
Here $L_n \left(x\right)$ denotes the Laguerre polynomial of order $n$.  
With the modified Wigner transformation in Eq.\eqref{eq:wignertransformation}, the Green function is not only gauge invariant but also invariant under center-of-mass spacetime translation.
It can be shown that, in the zero-magnetic-field limit, these reduce to

\begin{equation}
\begin{split}
G_0^{R/A}(\vec{p},\nu) &= \frac{1}{\nu-\frac{p^2}{2m} \pm i0^+},  \\
G_0^{K}(\vec{p},\nu) &= \-2\pi i \delta \left(\nu - \frac{p^2}{2m} \right)  \left( 1-2f(\nu)\right). 
\end{split}
\end{equation}

\subsection{Self-enegies}
For the self-energies, we consider an electron-impurity interaction.
In a real material, disorders are inevitably present. We model the interaction between electrons and impurities as 
\begin{equation}
S_{\text{imp}}[\psi^\dagger,\psi] = \int d\underline{x}_1  \sum_{a=1}^{2} \psi^\dagger_{a}(\underline{x}_1)  V_{\text{dis}}(\vec{r}_1) \psi_{a}(\underline{x}_1).
\end{equation}
For simplicity, we only consider the delta-correlated disorder potential $V_{\text{dis}}(\vec{r})$ of zero mean, \textit{i.e.},

\begin{equation}
	\langle V_{\text{dis}}(\vec{r})\rangle = 0 \hspace{0.2cm} \text{and} \hspace{0.2cm} \langle V_{\text{dis}}(\vec{r})V_{\text{dis}}(\vec{r}')\rangle = \gamma^2 \delta\left(\vec{r}-\vec{r}'\right).
\end{equation}
A generalization is straightforward. 

The self-energy for the electron-impurity interaction is evaluated within the self-consistent Born approximation (SCBA)~\cite{tsuneya1974I,tsuneya1974IV,Laikhtman1994}. Let us mention in passing here that this approximation can be realized as a stationary configuration of the non-linear sigma model for a disordered 2DEG \cite{KamenevAnton1999,KamenevLevchenko2009,Kamenev2011}. 

We find that, after averaging over the impurity ensemble, the self-energy is given by
\begin{equation}
	\hat{\Sigma}_{\text{imp}}(\vec{r},t,\vec{r}',t') =  \gamma^2 \delta(\vec{r}-\vec{r}') \hat{G}(\vec{r},t,\vec{r}',t').
\end{equation}
Its Wigner representation reads
\begin{equation}
\hat{\tilde{\Sigma}}_{\text{imp}}(\vec{p},\omega) = 2\pi \gamma^2 \int \frac{d\vec{k}}{(2\pi)^2} \frac{d\nu}{2\pi}~ \delta(\omega-\nu) \hat{\tilde{G}}(\vec{k},\nu).
\end{equation}
As the system is translationally invariant after disorder averaging, for brevity, we omit their space-time arguments.
The frequency integral can be done with the help of the delta function. 

First, we consider the retarded component of the self-energy. This gives
\begin{equation}
\label{eq:disorderselfenergy}
\tilde{\Sigma}^R_{\text{imp}}(\vec{p},\omega) = \gamma^2 \int \frac{d\vec{k}}{(2\pi)^2} \tilde{G}^R(\vec{k},\omega).
\end{equation}
It turns out that the self-consistent Green function remains diagonal in the Landau-level space. The reason is that

\begin{equation}
\begin{split}
\tilde{\Sigma}^R_{\text{imp}}(\vec{p},\omega) &= \gamma^2\int \frac{d\vec{k}}{(2\pi)^2} \sum_{n} 2(-1)^n e^{-\frac{k^2}{B}} L_n\left(\frac{2k^2}{B}\right)G^R_n(\omega), \\
&= \frac{\gamma^2}{2\pi/B} \sum_{n \ge 0}G^R_n(\omega).
\end{split}
\end{equation}
Expressing the self-energy in the Landau-level basis, one obtains
\begin{equation}
\begin{split}
&\sum_{m} 2(-1)^m e^{-\frac{p^2}{B}} L_m\left(\frac{2p^2}{B}\right) \Sigma^R_m(\omega)  \\ &= \underbrace{\sum_{m} 2(-1)^m e^{-\frac{p^2}{B}} L_m\left(\frac{2p^2}{B}\right)}_{=1}\frac{\gamma^2}{2\pi/B} \sum_{n \ge 0}G^R_n(\omega).
\end{split}
\end{equation}
This implies that 
\begin{equation}
\label{eq:scselfenergy}
\Sigma^R_m(\omega) = \frac{\gamma^2}{2\pi/B}\sum_{n \ge 0}G^R_n(\omega).
\end{equation}
Here $\Sigma^R_m$ represents the self-energy contribution to the Landau level $m$. The self-consistent Green function above takes the form
\begin{equation}
G^R_n(\omega) = \frac{1}{\omega - \epsilon_n - \Sigma^R_n({\omega})} .
\end{equation}

From Eq.\eqref{eq:scselfenergy}, we observe that every Landau level contributes to the self-energy regardless of the electronic occupancy. The summation extends over all Landau levels. For technical simplicity, we extend the summation over from $n=-\infty$ to $n=\infty$. With this assumption, there exist a self-consistent solution $\Sigma(\omega)$ defined by $\Sigma^R_m(\omega) = \Sigma(\omega-m\omega_c)$. It is a periodic function with period $\omega_c$. In other words, $\Sigma(\omega)=\Sigma(\omega+n\omega_c)$ for any integer $n$. Thus, it is sufficient to solve the self-consistent equation at frequencies around the lowest Landau level, i.e., $0 \le \omega  \le \omega_c$.

\begin{equation}
\Sigma(\omega) = \frac{\gamma^2}{2\pi/B} \sum_{n=-\infty}^{\infty} \frac{1}{\omega-(n+1/2)\omega_c-\Sigma(\omega)}.
\end{equation}

The summation can be evaluated exactly with the help of a contour integral \cite{tsuneya1974IV}. This gives
\begin{equation}
\sigma(z)  \tan\left(\pi\left(z-1/2 -\sigma(z)\right)\right) = \frac{(\gamma/\omega_c)^2}{2/B}.
\end{equation}
Here the summation was carried out in the dimensionless units, i.e., $z=\omega/\omega_c$ and $\sigma = \Sigma/\omega_c$.
In the strong-field limit, the tangent function can be expand to the lowest order. We obtain a quadratic equation for the self-energy reading as
\begin{equation}
\sigma(z)\left( z-1/2-\sigma(z)\right) = \frac{(\gamma/\omega_c)^2}{2\pi/B}
\end{equation}
of which the solution can be easily obtained. After restoring the dimensionful parameter $\omega_c$, the self-energy reads as
\begin{equation}
\Sigma(\omega) = \frac{(\omega-\omega_c/2) + \sqrt{(\omega-\omega_c/2)^2-\frac{2\gamma^2}{\pi/B}}}{2}.
\end{equation}


\section{Quantum Hall Physics in a disordered 2DEG}
\label{sec:results}

In this section, we derive and solve the quantum kinetic equation for a disordered 2DEG in the quantum hall regime. We calculate both longitudinal and Hall conductivities.

\subsection{Kinetic equation}

As an example, consider a gas of non-relativistic electrons exposed to an external magnetic field of arbitrary strength and a weak electric field. According to the Wigner transform \eqref{eq:wignertransformation}, the inverse propagator \eqref{eq:inversepropoagator} is mapped to 
\begin{equation}
\tilde{G}^{-1}_0(\omega,\vec{p}) = \omega-\frac{\vec{p}^2}{2m}.
\end{equation}
This allows us to evaluate the streaming terms. We find that
\begin{widetext}
\begin{equation}
\left[G^{-1}_0~ \overset{\circ}{,} ~G^{K} \right] \xrightarrow{\text{WT}}   i \left[ \partial_{T} +\frac{\vec{p}}{m}\cdot \partial_{\vec{R}}   
 - \vec{E}\cdot\frac{\vec{p}}{m} \partial_{\omega} - \vec{E}\cdot\partial_{\vec{p}} -\left\{ \frac{\vec{p}}{m} \times \vec{B}\right\} \cdot  \partial_{\vec{p}} \right] \tilde{G}^K.
\end{equation}
\end{widetext}
Note that, there is a frequency derivative term $-\vec{E}\cdot\frac{\vec{p}}{m} \partial_{\omega}\tilde{G}^K$  in addition to the standard Liouville's operator. It turns out that this term is important in realising a series of peaks in the longitudinal electrical conductivity of quantum Hall state. 

Together with the self-energies, we obtain the quantum kinetic equation reading as

\begin{widetext}
\begin{equation}
\label{eq:fermionkineticequation}
\begin{split}
 &i \left[  \partial_{T} +\frac{\vec{p}}{m}\cdot \partial_{\vec{R}}   
 - \vec{E}\cdot\frac{\vec{p}}{m} \partial_{\omega} - \vec{E}\cdot\partial_{\vec{p}} -\left( \frac{\vec{p}}{m} \times \vec{B}\right) \cdot  \partial_{\vec{p}} \right] \tilde{G}^K= \\ &\hspace{4cm}=  \gamma^2 \int \frac{d\vec{k}}{(2\pi)^2} \left( \left[G^R(\vec{k},\nu)-G^A(\vec{k},\nu)\right]G^K(\vec{p},\nu) - \left[G^R(\vec{p},\nu)-G^A(\vec{p},\nu)\right]G^K(\vec{k},\nu)\right).
 \end{split}
\end{equation}
\end{widetext}
This is one of the main results of our paper.
The transport equation describes time evolution of the Keldysh Green function of four variables $(\vec{p},\vec{R},\nu,T)$. It treats the energy $\nu$ and momentum $\vec{p}$ as independent variables. As such, it is applicable beyond the quasiparticle approximation. 

It should be emphasize that the Lorentz force term, $\left( \frac{\vec{p}}{m} \times \vec{B}\right) \cdot  \partial_{\vec{p}} \tilde{G}^K$, does not restrict the validity of the equation to the weak magnetic field limit. This is only a result of the gradient-expansion approximation. The system changes slowly in coordinate space and only the first order in gradient is kept. In fact, with the help of the Landau level basis, the Keldysh Green function can be evaluated to an arbitrary order of  magnetic field strength.  The equation is valid to all order in magnetic field.


To evaluate the dc transport coefficients, we will be looking for a homogeneous and steady-state solution of Eq.\eqref{eq:fermionkineticequation}. Thus, the terms with space and time derivatives are neglected and the space and time variables of the Green functions will be dropped. This gives
\begin{widetext}
\begin{equation}
\label{eq:QKE}
\begin{split}
	&i \left[ -\vec{E}\cdot\frac{\vec{p}}{m}\partial_\nu- \left(\vec{E}+\frac{\vec{p}}{m} \times \vec{B}\right) \cdot \partial_{\vec{p}} \right] G^K(\vec{p},\nu) \\ &\hspace{4cm}=  \gamma^2 \int \frac{d\vec{k}}{(2\pi)^2} \left( \left[G^R(\vec{k},\nu)-G^A(\vec{k},\nu)\right]G^K(\vec{p},\nu) - \left[G^R(\vec{p},\nu)-G^A(\vec{p},\nu)\right]G^K(\vec{k},\nu)\right).
\end{split}
\end{equation}
\end{widetext}

It should be noted that this equation is valid for arbitrary magnetic field strength.  In fact, the Green function is carried out in the Landau level basis which is the exact eigenstates for electrons subjected to a uniform magnetic field of any intensity.

\subsection{Electrical conductivities}


When a 2D electron gas is exposed to a perpendicular strong magnetic fields at low temperature, its Hall conductivity $\sigma_{yx}$ is given by a multiple of the fundamental constants $e^2/h$. The  system exhibits plateaus in $\sigma_{yx}$ together with a series of peaks in the longitudinal conductivity $\sigma_{xx}$. This is known as the quantum Hall effect (QHE).  In this section, we will solve the quantum kinetic equation within the linear response and seek a solution for the quantum Hall conditions. To this end, we parametrize
\begin{equation}
\label{eq:parametrizeKeldyshGreenfunction}
	G^K(\vec{p},\nu) = G_{\text{eq}}^{K}(\vec{p},\nu) + \delta G^K(\vec{p},\nu).
\end{equation}
The equilibrium Keldysh Green function $G^K_{\text{eq}}(\vec{p},\nu)$ is defined in \label{eq:GreenfunctioninLandaulevelbasis} with $f$ being the Fermi-Dirac function. It nullifies the collision integral as well as the driving term. Since we are interested in the linear response to the electric field, the deviation from the equilibrium solution assumes the form

\begin{equation}
\begin{split}
\label{eq:parametrization}
	\delta G^{K}(\vec{p},\nu) &= \chi_{\parallel} \vec{E}\cdot\left(  \partial_{\vec{p}} +\frac{\vec{p}}{m}\partial_\nu\right)G_{\text{eq}}^K(\vec{p},\nu) \\&+ \chi_\perp \left(\vec{E} \times \hat{B} \right) \cdot \left(  \partial_{\vec{p}} +\frac{\vec{p}}{m}\partial_\nu\right) G^K_{\text{eq}}(\vec{p},\nu),
\end{split}
\end{equation}
Here $\chi_{\parallel}$ and $\chi_{\perp}$ are to be determined from the quantum kinetic equation. For simplicity, we assume that they are momentum and energy independent. 
In principle, this constraint may be relaxed. 

For a notational convenient, we denote $\partial_{\vec{p}}G^K_{\text{eq}}(\vec{p},\nu) = \vec{p} ~\tilde{G}^K_{\text{eq}}(p,\nu)$. We proceed to the linearization of Eq.\eqref{eq:QKE} by inserting the parametrization in Eq.\eqref{eq:parametrization} and keep terms of linear order in $\vec{E}$. 
After a few algebraic manipulations, we obtain the linearized quantum kinetic equation reading as
\begin{widetext}
\begin{equation}
\begin{split}
	&-\vec{E}\cdot \vec{p} \left(\tilde{G}^K_{\text{eq}}(p,\nu)+\frac{1}{m}\partial_\nu G^K_{\text{eq}}(\vec{p},\nu)\right)   \\ &\hspace{0.5cm}= -\left(\vec{E} \times \vec{B}\right) \cdot \frac{\vec{p}}{m} \left( \tilde{G}_{\text{eq}}^K(p,\nu) +\frac{1}{m} \partial_\nu G^K_{\text{eq}}(\vec{p},\nu) \right) \chi_{\parallel}  + B\vec{E} \cdot \frac{\vec{p}}{m}  \left( \tilde{G}_{\text{eq}}^K(p,\nu) +\frac{1}{m} \partial_\nu G^K_{\text{eq}}(\vec{p},\nu) \right) \chi_{\perp} \\ & \hspace{0.5cm}~~~~-i\gamma^2 \int \frac{d\vec{k}}{(2\pi)^2}  \Big[G_{\text{eq}}^R(\vec{k},\nu)-G_{\text{eq}}^A(\vec{k},\nu)\Big] \left[ \chi_{\parallel} \vec{E}\cdot\left(  \partial_{\vec{p}} +\frac{\vec{p}}{m}\partial_\nu\right)G_{\text{eq}}^K(\vec{p},\nu) + \chi_\perp \left(\vec{E} \times \hat{B} \right) \cdot \left(  \partial_{\vec{p}} +\frac{\vec{p}}{m}\partial_\nu\right) G^K_{\text{eq}}(\vec{p},\nu) \right].
\end{split}
\end{equation}
\end{widetext}
The equation describes the change of the Keldysh Green function along $\vec{E}\cdot\vec{p}$ and $\left(\vec{E}\times \hat{B}\right)\cdot\vec{p}$ directions. Each component reads as
\begin{widetext} 
\begin{equation}
\label{eq:linearizedQKEparallel}
\begin{split}
	- \left(\tilde{G}^K_{\text{eq}}(p,\nu)+\frac{1}{m}\partial_\nu G^K_{\text{eq}}(\vec{p},\nu)\right)   &= \frac{B}{m}  \left( \tilde{G}_{\text{eq}}^K(p,\nu) +\frac{1}{m} \partial_\nu G^K_{\text{eq}}(\vec{p},\nu) \right) \chi_{\perp} \\ & \hspace{0.5cm}-i\gamma^2 \int \frac{d\vec{k}}{(2\pi)^2}  \Big[G_{\text{eq}}^R(\vec{k},\nu)-G_{\text{eq}}^A(\vec{k},\nu)\Big] \left[  \tilde{G}^K_{\text{eq}}(p,\nu) +\frac{1}{m}\partial_\nu G_{\text{eq}}^K(\vec{p},\nu) \right] \chi_{\parallel}, 
\end{split}
\end{equation}
and
\begin{equation}
\label{eq:linearizedQKEperp}
\begin{split}
	0= -\frac{B}{m} \left( \tilde{G}_{\text{eq}}^K(p,\nu) +\frac{1}{m} \partial_\nu G^K_{\text{eq}}(\vec{p},\nu) \right) \chi_{\parallel} -i\gamma^2 \int \frac{d\vec{k}}{(2\pi)^2}  \Big[G_{\text{eq}}^R(\vec{k},\nu)-G_{\text{eq}}^A(\vec{k},\nu)\Big] \left[   \tilde{G}^K_{\text{eq}}(p,\nu) +\frac{1}{m}\partial_\nu G_{\text{eq}}^K(\vec{p},\nu) \right] \chi_{\perp} .
\end{split}
\end{equation}
\end{widetext}

The equations can be casted into a matrix equation reading as
\begin{equation}
\begin{pmatrix}
-1 \\ 0
\end{pmatrix} = 
\begin{pmatrix}
D(\nu) & \omega_c \\ -\omega_c & D(\nu)
\end{pmatrix} 
\begin{pmatrix}
\chi_{\parallel} \\ \chi_{\perp}
\end{pmatrix}. 
\end{equation}
Here, for a notational convenience, we introduce a shorthand
\begin{equation}
\begin{split}
D(\nu) &= -i\gamma^2 \int \frac{d\vec{k}}{(2\pi)^2} \left[ G_{\text{eq}}^R(\vec{k},\nu)-G^{A}_{\text{eq}}(\vec{k},\nu)\right] \\&= -i\frac{\gamma^2}{2\pi/B} \sum_n \left[ G^R_n(\nu) - G^A_n(\nu)  \right] \\&=2 \sum_{n} \Im \Sigma_n^R(\nu).
\end{split}
\end{equation}
The solutions is therefore given by
\begin{equation}
\begin{pmatrix}
\chi_{\parallel} \\ \chi_{\perp}
\end{pmatrix}  = -\frac{1}{D^2(\nu)+\omega_c^2}
\begin{pmatrix}
D(\nu) \\ \omega_c
\end{pmatrix}.
\end{equation}

To calculate the electrical conductivity, we need to evaluate the current density to linear order in the electric field. The current density for 2DEG is given by
\begin{equation}
\vec{j}(\vec{r}_1,t_1) = \frac{i}{2} \frac{1}{m} \left(-i\partial_{\vec{r}_1} + \vec{A}(\vec{r}_1,t_1) \right) G^K(\vec{r}_1,t_1,\vec{r}_2,t_2) \Big|_{\substack{\vec{r}_1=\vec{r}_2 \\ t_1=t_2}}.
\end{equation}
In the Wigner space, it reads
\begin{equation}
\begin{split}
\vec{j}(\vec{r}_1,t_1) 
&=\frac{i}{2} \int \frac{d\vec{p}}{(2\pi)^2}\frac{d\omega}{2\pi}  \frac{\vec{p}}{m} G^K\left(\vec{p},\vec{r}_1,\omega,t_1\right).
\end{split} 
\end{equation}
As the system remains homogeneous and static, the current appears independent of space-time coordinates. Inserting the Keldysh Green function in Eq.\eqref{eq:parametrizeKeldyshGreenfunction} to the current density. We can extract the conductivities and find that
\begin{widetext}
\begin{equation}
\sigma_{xx} =  -\frac{i}{2}\int \frac{d\vec{p}}{(2\pi)^2} \frac{d\nu}{2\pi} \frac{p_x^2}{m} \left(  \tilde{G}_{\text{eq}}^K(p,\nu) +\frac{1}{m}\partial_\nu G_{\text{eq}}^K(\vec{p},\nu) \right)\frac{D(\nu)}{D^2(\nu)+\omega_c^2},
\end{equation}
\begin{equation}
\sigma_{yx} =  \frac{i}{2}\int \frac{d\vec{p}}{(2\pi)^2} \frac{d\nu}{2\pi} \frac{p_y^2}{m} \left(  \tilde{G}_{\text{eq}}^K(p,\nu) +\frac{1}{m}\partial_\nu G_{\text{eq}}^K(\vec{p},\nu) \right) \frac{\omega_c}{D^2(\nu)+\omega_c^2} ,
\end{equation}
\end{widetext}
These expressions constitute the main results of this work.

\begin{figure}[t]
		\includegraphics[width=0.3\textwidth]{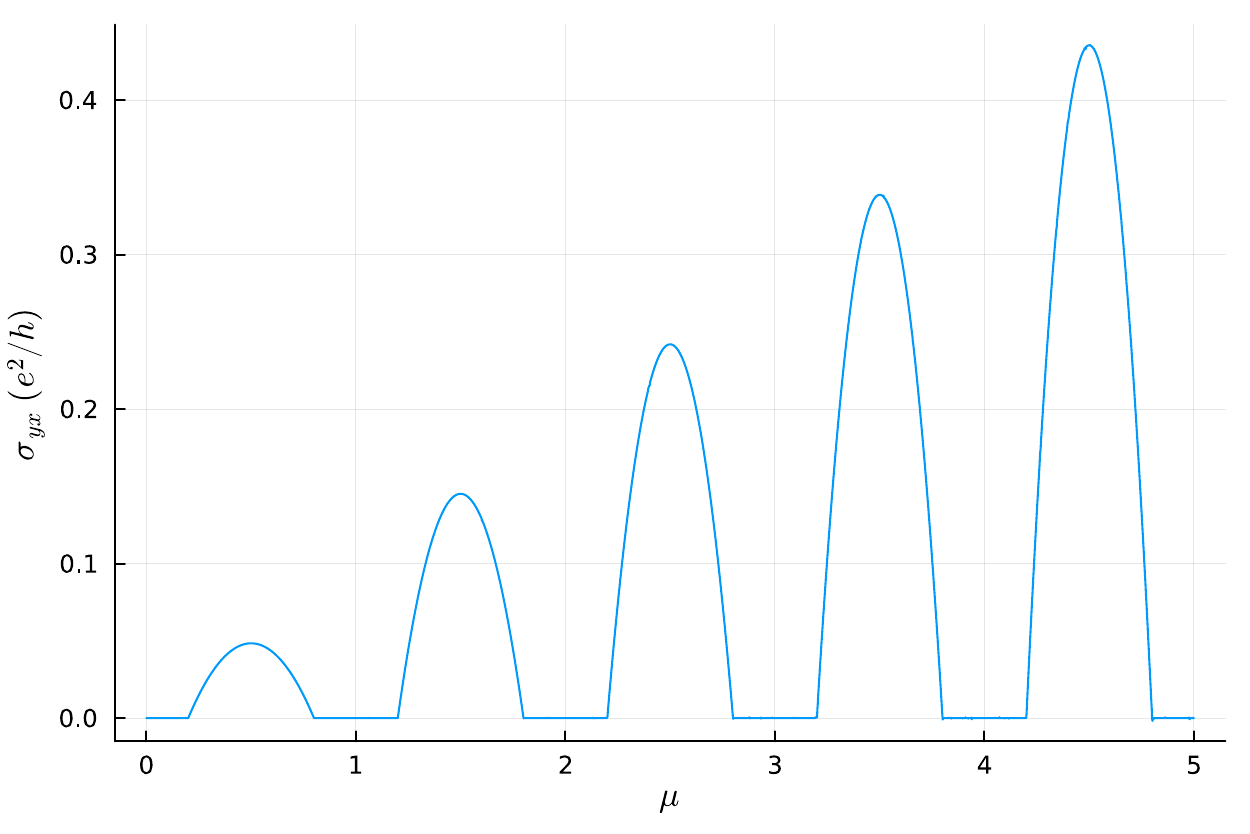}
		\caption{The longitudinal conductivity $\sigma_{xx}$ as a function of the chemical potential at zero temperature}
		\label{fig:sigmaxx}
\end{figure}

\begin{figure}[t]
		\includegraphics[width=0.3\textwidth]{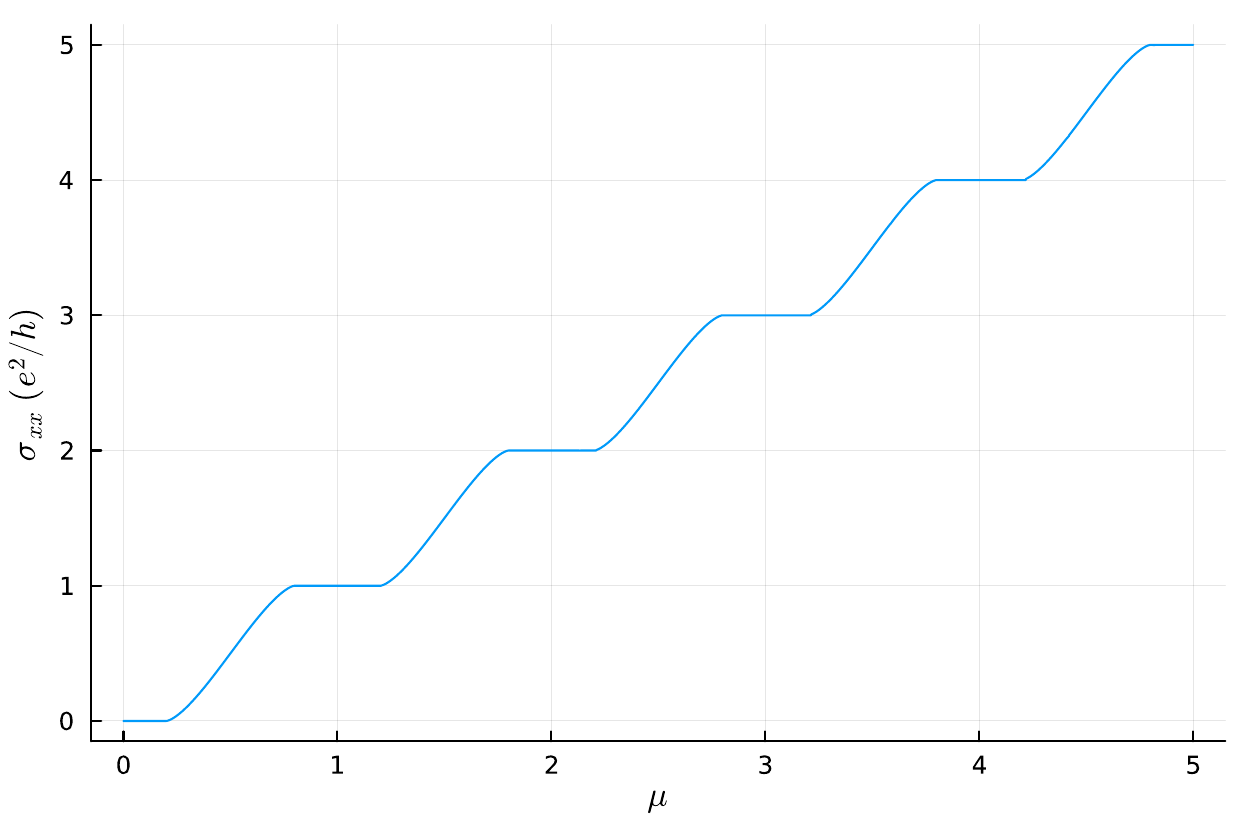}
		\caption{The Hall conductivity $\sigma_{yx}$ as a function of the chemical potential at zero temperature }
		\label{fig:sigmayx}
\end{figure}

The quantum Hall regime occurs in the strong magnetic field limit, \textit{i.e.}, $D(\nu) \ll \omega_c$. For simplicity, we will evaluate the conductivities at zero temperature. We find that the Hall conductivity is quantized in units of $e^2/\hbar$ as

\begin{equation}
\sigma_{yx} 
= \frac{e^2}{h}\sum_{n}\int_0^\mu \frac{d\nu}{\pi} \Im G_n^R(\omega) + \mathcal{O} \left( \frac{D(\nu)}{\omega_c}\right). 
\end{equation}
Here the integral $\sum_{n}\int_0^\mu \frac{d\nu}{\pi} \Im G_n^R(\omega)  \in \mathbb{Z}^{+} \cup \{0\} $ counts the number of Landau levels below the chemical potential $\mu$.
In addition, we evaluate the diagonal conductivity. We find that

\begin{equation}
\begin{split}
\sigma_{xx}  =& \frac{e^2}{h} \left[-\frac{1}{\pi}  \sum_n \left( n+\frac{1}{2}\right) \Im  G^R_n(\mu) D(\mu) \right. \nonumber \\ &\left. + \frac{1}{\pi} \int^{\mu}_{0} d\nu\sum_{n} \left( n+\frac{1}{2}\right)D(\nu) \partial_{\nu}\Im G_n^R(\nu)  \right]
\end{split}
\end{equation}
As we show in Figs. \ref{fig:sigmaxx} and \ref{fig:sigmayx}, the solution of the quantum kinetic equation gives the quantum Hall plateau in the Hall conductivity together with a series of peaks in $\sigma_{xx}$. 

\section{Summary and Outlooks}
\label{sec:conclusion}

In this paper, we present a systematic derivation of the quantum kinetic equation from the Keldysh quantum field theory using the Landau-level basis. As an application of our quantum transport equation, we calculate the electrical conductivity of a 2DEG in the quantum hall regime in the presence of impurities. We find that, as a result of the discreteness of the Landau levels, a solution to our quantum kinetic equation shows the plateaus of Hall conductivity accompanied by a series of peaks in the longitudinal conductivity.  

It is interesting to study the effects of other scattering mechanisms, for example, due to electron-phonon \cite{MacDonald1991,Pound2011} and electron-electron interactions \cite{silin1958} in electrical conductivities as well as other transport coefficients such as thermal conductivity and viscosity. We leave this for future investigations not only for the 2DEG but also other system like Dirac materials.

\section{Acknowledgement}
KP acknowledges discussions and former collaborations with Lars Fritz, Henk Stoof, and Tim Ludwig, Tobias Meng. This research project was supported by the Faculty of Science, Mahidol University.
\bibliographystyle{apsrev4-2}
\bibliography{references.bib}

\end{document}